\documentclass[final,5p,times,twocolumn]{elsarticle}

\usepackage{graphicx}
\usepackage{float} 
\usepackage{gensymb}
\usepackage{hyperref}
\hypersetup{hidelinks}
\usepackage{amssymb}

\journal{Nuclear Inst. and Methods in Physics Research, B}

\begin{document}

\begin{frontmatter}



\title{RAPTOR: a new collinear laser ionization spectroscopy and laser-radiofrequency double-resonance experiment at the IGISOL facility}


\author[inst1]{S. Kujanpää\corref{cor1}}

\affiliation[inst1]{organization={Department of Physics,  Accelerator laboratory, University of Jyväskylä},
            city={Jyväskylä},
            postcode={FI-40014 },
            country={Finland}}
        
\author[inst1]{A. Raggio}
\author[inst1,inst3]{R. P. de Groote}

\author[inst3,inst2]{M. Athanasakis-Kaklamanakis}
\author[inst5,jgu,inst6]{M. Block}
\author[inst3]{A. Candiello}
\author[inst1]{W. Gins}
\author[inst4]{\'{A}. Koszor\'{u}s}
\author[inst1]{I. D. Moore}
\author[inst1]{M. Reponen}
\author[inst5,jgu]{J. Warbinek}

\cortext[cor1]{Corresponding author: sopikuja@jyu.fi}

\affiliation[inst3]{organization={Instituut voor Kern- en Stralingsfysica, KU Leuven},
            postcode={B-3001 Leuven}, 
            country={Belgium}}
\affiliation[inst2]{organization={Experimental Physics Department, CERN},
            postcode={CH-1211 Geneva 23},
            country={Switzerland}}

\affiliation[inst5]{organization={GSI Helmholtzzentrum fur Schwerionenforschung GmbH}, postcode={Darmstadt, 64291}, country = {Germany}}

\affiliation[jgu]{  organization={Department Chemie - Standort TRIGA, Johannes Gutenberg - Universität},
                    city={55099 Mainz},
                    country={Germany}}
\affiliation[inst6]{organization={Helmholtz-Institut Mainz}, postcode={Mainz, 55128}, country = {Germany}}
\affiliation[inst4]{organization={Department of Physics, University of Liverpool},
            postcode={Liverpool L69 7ZE}, 
            country={United Kingdom}}

\begin{abstract}

RAPTOR, Resonance ionization spectroscopy And Purification Traps for Optimized spectRoscopy, is a new collinear resonance ionization spectroscopy device constructed at the Ion Guide Isotope Separator On-Line (IGISOL) facility at the University of Jyväskylä, Finland. By operating at beam energies of under 10\,keV, the footprint of the experiment is reduced compared to more traditional collinear laser spectroscopy beamlines. In addition, RAPTOR is coupled to the JYFLTRAP Penning trap mass spectrometer, opening a window to laser-assisted nuclear-state selective purification, serving not only the mass measurement program, but also supporting post-trap decay spectroscopy experiments. Finally, the low-energy ion beams used for RAPTOR will enable high-precision laser-radiofrequency double-resonance experiments, resulting in spectroscopy with linewidths below 1 MHz. In this contribution, the technical layout of RAPTOR and a selection of ion-beam optical simulations for the device are presented, along with a discussion of the current status of the commissioning experiments.

\end{abstract}




\end{frontmatter}



\begin{figure*}[t]
    \centering
    \includegraphics[width=1\textwidth]{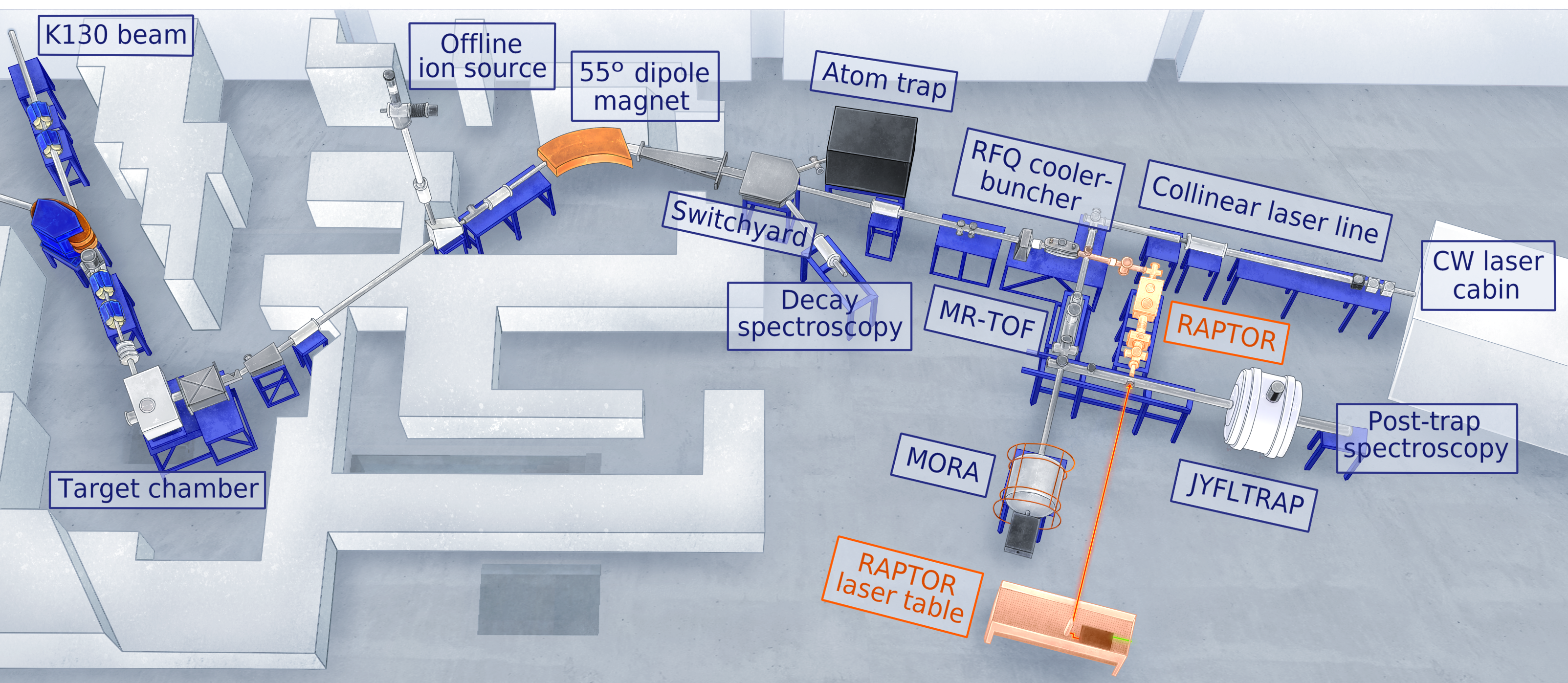}
    \caption{Illustration of the current IGISOL facility. Primary beams from the K130 heavy-ion cyclotron are guided to the target chamber where secondary beams are produced in various nuclear reactions. Offline beams are produced using a discharge ion source that can be mounted inside the target chamber or an offline ion source located on the second floor. A dipole magnet is used to mass-separate the ion beam, after which it is guided to an RFQ cooler-buncher. The bunched and cooled beam is delivered to the RAPTOR beamline, highlighted in orange. An optical table dedicated to RAPTOR as well as MORA is indicated. See text for more details.}
    \label{fig:igi}
\end{figure*}

\section{Introduction} \label{sec:intro}

Laser spectroscopy experiments have proven to be exceptionally well-suited to test our understanding of the atomic nucleus~\cite{YANG2023104005}. They provide nuclear-model independent access to nuclear electromagnetic moments, spins, and changes in the mean-squared nuclear charge radii. The Collinear Resonance Ionization Spectroscopy (CRIS) technique with bunched beams was pioneered in the past decade at ISOLDE-CERN (see e.g., \cite{cris, cris1,cris2}). It exploits the high selectivity of resonance laser ionization, high ion detection efficiency, and the reduced Doppler broadening offered by fast beams. To expand the optical spectroscopy capabilities of radioactive isotopes, a new CRIS device has been constructed at the IGISOL facility \cite{moore2014igisol} 
located in the Accelerator Laboratory of University of Jyväskylä (JYFL-ACCLAB) \cite{acclab}, Finland. In contrast to other collinear laser spectroscopy experiments, RAPTOR (Resonance ionization spectroscopy And Purification Traps for Optimized spectRoscopy) is designed to operate at beam energies below 10\,keV. Thus, the neutralization into the states of interest can be optimized due to the increased selectivity of charge-exchange reactions at lower beam energies \cite{vernon2019simulation}. These lower energy ion beams will also allow for  high-precision laser-radiofrequency double resonance experiments to be performed, resulting in spectroscopy with linewidths well below 1 MHz. This will enable measurements of e.g., hyperfine anomalies or other higher-order effects. In this article, we discuss the design of the beamline, ion-optical simulations of the beam transport and present the results of the first commissioning experiments.

\section{Description of the experiment}
\label{sec:exp}

\subsection{IGISOL facility}

At the IGISOL facility, illustrated in Fig. \ref{fig:igi}, ion beams are produced either with offline ion sources or on-line in nuclear reactions driven with primary beams from the K130 heavy-ion cyclotron. Further downstream from the target chamber and the vertical offline ion source line, a dipole sector magnet (mass resolving power $R  \approx 400$) mass-separates the species of interest towards the electrostatic switchyard where the ion beam can be directed to either a decay spectroscopy station, an atom trap, or the radio-frequency quadrupole cooler-buncher (RFQ) device. The ion bunches,  extracted from the RFQ at a 2-keV energy, are delivered to various experimental stations.  These include a fast-beams collinear laser spectroscopy line, RAPTOR, and a variety of ion trapping devices; an MR-TOF (Multi-Reflection Time-Of-Flight mass spectrometer), MORA \cite{mora} (Matter's Origin from the RadioActivity of trapped and oriented ions setup), and the double Penning trap JYFLTRAP \cite{jyfltrap}. RAPTOR is installed directly downstream from the RFQ and connects to the JYFLTRAP line. This allows for a unique combination of element-selective capabilities of resonance laser ionization with high-resolution mass spectrometry and/or mass purification for post-trap nuclear decay spectroscopy experiments. For laser spectroscopy, laser light can be produced in a number of areas of the facility, including the FURIOS (Fast Universal Resonant laser IOn Source) \cite{furios} pulsed lasers located on the second floor of the facility , a continuous-wave (CW) laser cabin  and a smaller shielded laser table which produces pulsed laser beams for RAPTOR and MORA.
 

\begin{figure*}[t]
    \centering
    \includegraphics[width=1\textwidth]{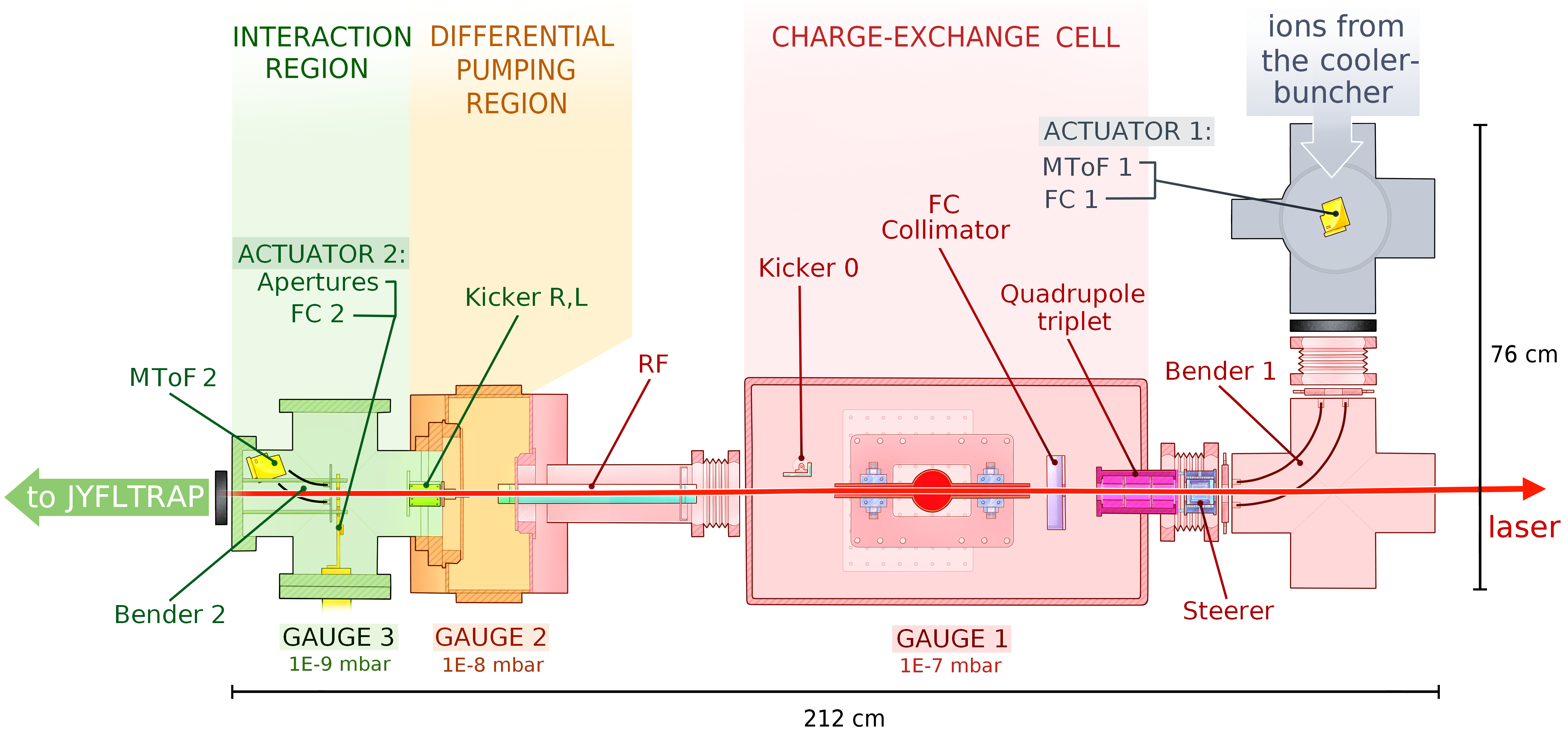}
    \caption{Illustration of the RAPTOR beamline. The bunched ion beam arrives from the IGISOL RFQ cooler-buncher (top right). Magnetof detectors (MToF) and Faraday cups (FC) are used for beam diagnostics. The charge-exchange cell (CEC) converts ions into atoms as they interact with a hot neutralizer vapour. Kicker electrodes are used to remove non-neutralized ions. The radiofrequency (RF) electrode can be used for laser-radiofrequency double-resonance spectroscopy in the future. The different pressure regions are indicated in red, orange, and green. More details in the text.}
    \label{fig:raptor}
\end{figure*}

\subsection{RAPTOR beamline}

A schematic layout of the RAPTOR beamline is presented in Fig. \ref{fig:raptor}, illustrating the ion-optical elements and general dimensions of the device. Bunched ions produced by the RFQ are guided and focused with steerer electrodes and an einzel lens towards a set of extractor electrodes (not pictured), which adjust the focus of the ion beam as it is accelerated to the RAPTOR beam energy. The potential of the RAPTOR beamline is defined relative to the RFQ platform voltage and it is set with a -10 kV Spellman power supply. Immediately after this acceleration, a magnetof (MToF) particle detector and a suppressed faraday cup (FC) are installed on a motorized remotely-controlled actuator to provide beam diagnostics, in a chamber which is separated from the rest of the RAPTOR beam line with a DN100CF gate valve. Next, the beam enters a double-focusing 90\textdegree \space deflector, based on the design in \cite{kreckel2010}, consisting of two parallel plates of different heights with a 90\textdegree \space curvature. The external plate has a hole covered with a metallic mesh in-line with RAPTOR main axis to allow laser light transmission. This bender guides the beam into the charge-exchange cell (CEC) chamber. The voltages required to bend the beam are approximately 1/4 of the beam energy. By applying an asymmetric voltage to the plates, the focus in the horizontal and vertical planes can be tuned. Before reaching the CEC, the beam passes through a pair of horizontal and vertical parallel steerer plates. A quadrupole triplet then refocuses the beam. An electrically-isolated collimator plate is located at the entrance of the CEC, acting as a collimator-faraday cup and a shield against alkaline vapour that may diffuse out of the CEC. Within the CEC, charge-exchange reactions between the ion beam and a vapour of a suitable alkaline element, such as potassium or sodium,  neutralizes the ions. The cell is electrically isolated from ground potential to allow for Doppler tuning of the atomic beam into resonance with the laser beams by applying an accelerating potential. After the CEC, a horizontal kicker electrode (kicker 0), can be used to remove the remaining ions from the neutralized beam. 

To reduce the rate of unwanted collisional ionization, the pressure in the atom-laser interaction region should be as good as possible. To shield the interaction region from the elevated pressures in the CEC chamber, typically $10^{-7}$ mbar when in operation, a differential pumping stage is used. The differential pumping chamber is separated from the charge-exchange chamber by a long (30 cm) coaxial radiofrequency (RF) electrode, similar to the design in \cite{vanhove}. This electrode serves as a conductance barrier ensuring good differential pumping. In the future, the electrode will be used for collinear laser-RF double-resonance measurements of radioactive isotopes. During the commissioning of RAPTOR, this electrode is grounded to avoid unwanted steering of the beam.  The laser-atom interaction chamber, shown in green in Fig. \ref{fig:raptor}, is located after the differential pumping barrier, and houses a set of horizontal kicker electrodes (kicker R, L). These are used to remove ions produced via collisions with the rest gas in the region between the CEC and the interaction region. Laser radiation is introduced in a counter-propagating geometry to the atomic beam. When all lasers are overlapped both spatially and temporally and are tuned into resonance with an atomic transition, the valence electron is excited in a stepwize process until the atom is ionized. The resonantly-produced ions are detected with an off-axis MToF detector, placed after a small bender electrode.To optimize the transport of the resonantly-produced ions and the laser light through RAPTOR, different sized apertures along with a suppressed faraday cup are mounted on a motorized actuator installed in the interaction region. The aperture ladder is directly in front of the bender to the MToF which ensures the spatial overlap of the laser and ion beams. 
Additional 90\degree \space bender electrodes are installed to deliver resonant ions from RAPTOR to JYFLTRAP. Furthermore, additional diagnostics, including an MToF detector, have been added on a 4-way cross on the opposite flange of the JYFLTRAP beam line facing RAPTOR in order to optimize the transmission efficiency and characterize the neutralisation efficiency reliably. A optical table with a pulsed laser system and optics dedicated to the transport of laser beams to RAPTOR (see Fig. \ref{fig:igi} hosts one pulsed titanium-sapphire (Ti:Sa) $Z$-cavity \cite{rilis} pumped by a 10-kHz Nd:YAG laser, and an injection-locked Ti:Sa cavity \cite{seed}. The latter is seeded with CW laser light fibered from a Ti:Sa Matisse ring laser located in the CW laser cabin. Due to the multiple excitation steps involved in the resonant ionization process, pulsed laser beams can also be delivered from the FURIOS.

\section{Ion beam simulations}
\label{sec:sim}

\begin{figure}[t]
    \centering
    \includegraphics[width=1\columnwidth]{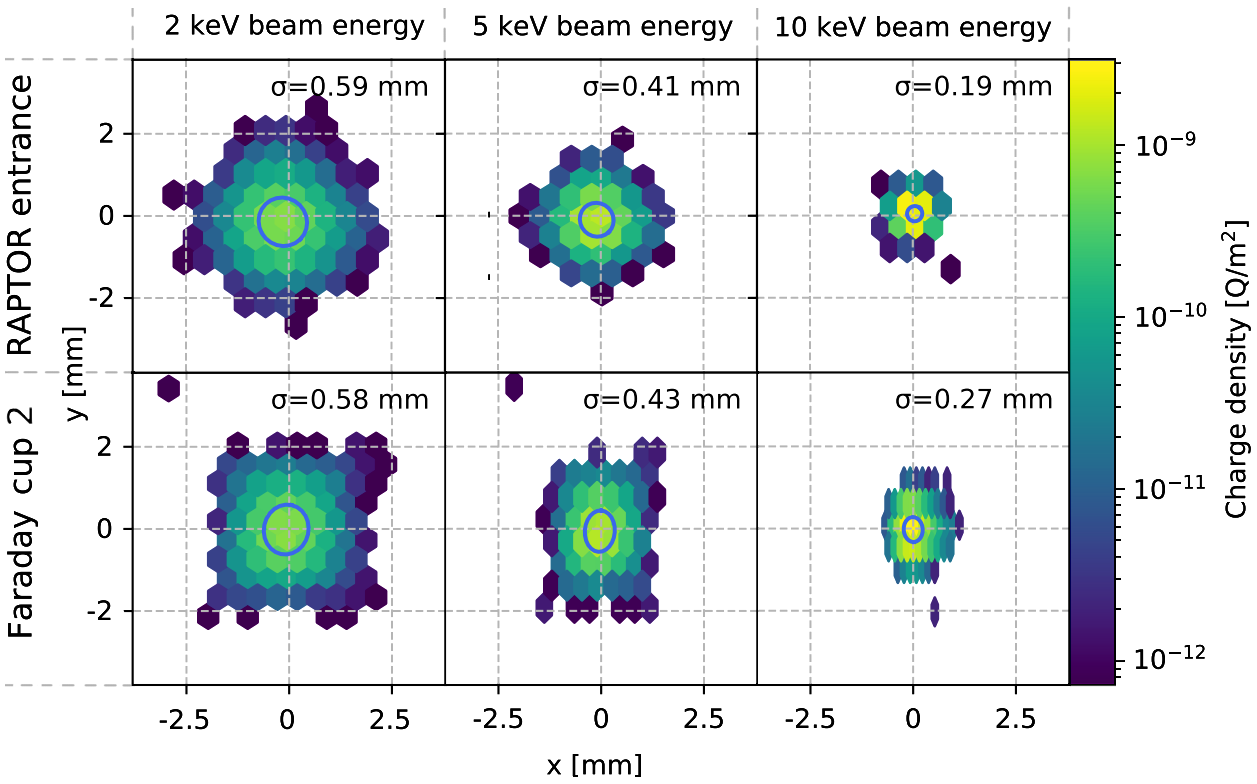}
    \caption{Simulated beam spots for the RAPTOR beamline at the entrance and FC 2 location for different ion beam energies. The calculated 1-$\sigma$ radius is provided for comparison.}
    \label{fig:spots}
\end{figure}

Ion beam transport through the RAPTOR beamline was simulated using two different methods. The first part of the transport was done with IBSimu \cite{ibsimu}, to model the extraction out of the RFQ and injection into the RAPTOR beamline, up to FC 1 in Fig.~\ref{fig:raptor}. The resulting spot on FC1 was optimized to be as parallel as possible in the simulations. After FC1, simulations were performed using the PIOL library, a software package under development at the University of Jyväskylä. The transport matrix formalism used here allows for the simulation of more complex beam lines without excessive computational requirements when assuming standard ion-optical elements.
Fig. ~\ref{fig:spots}~shows the simulated shape of the ion beam for 2, 5 and 10-keV beam energies at the RAPTOR entrance as well as FC 2 located in the interaction region. The beam spot sizes can be maintained throughout the beamline. Higher beam energies result in smaller beam spots due to the lower geometric emittance at higher energies. For all beam energies, sufficiently small spot sizes are obtained, indicating that a good geometrical overlap is possible with the counter-propagating laser beams.

\section{Commissioning results}
\label{sec:results}

The first RAPTOR commissioning experiments explored tuning of beamline's ion optical elements using various stable ion beams produced with the IGISOL offline ion sources. The voltages required on the ion-beam optics to transport beam through RAPTOR were found to be within 20\% of the simulated values. The transmission efficiency from FC 1 to FC 2 was determined to be around 60\% for a 5-keV beam energy and it remains relatively unchanged for other beam energies. The CEC was also tested for the first time; neutralization efficiencies of a few 10\% were estimated for Cu, Ag and Sn beams using the FC. More precise neutralization efficiency estimates will require the on-axis MToF detector which has been recently added after RAPTOR.  

\begin{figure}[h]
    \centering
    \includegraphics[width=1\columnwidth]{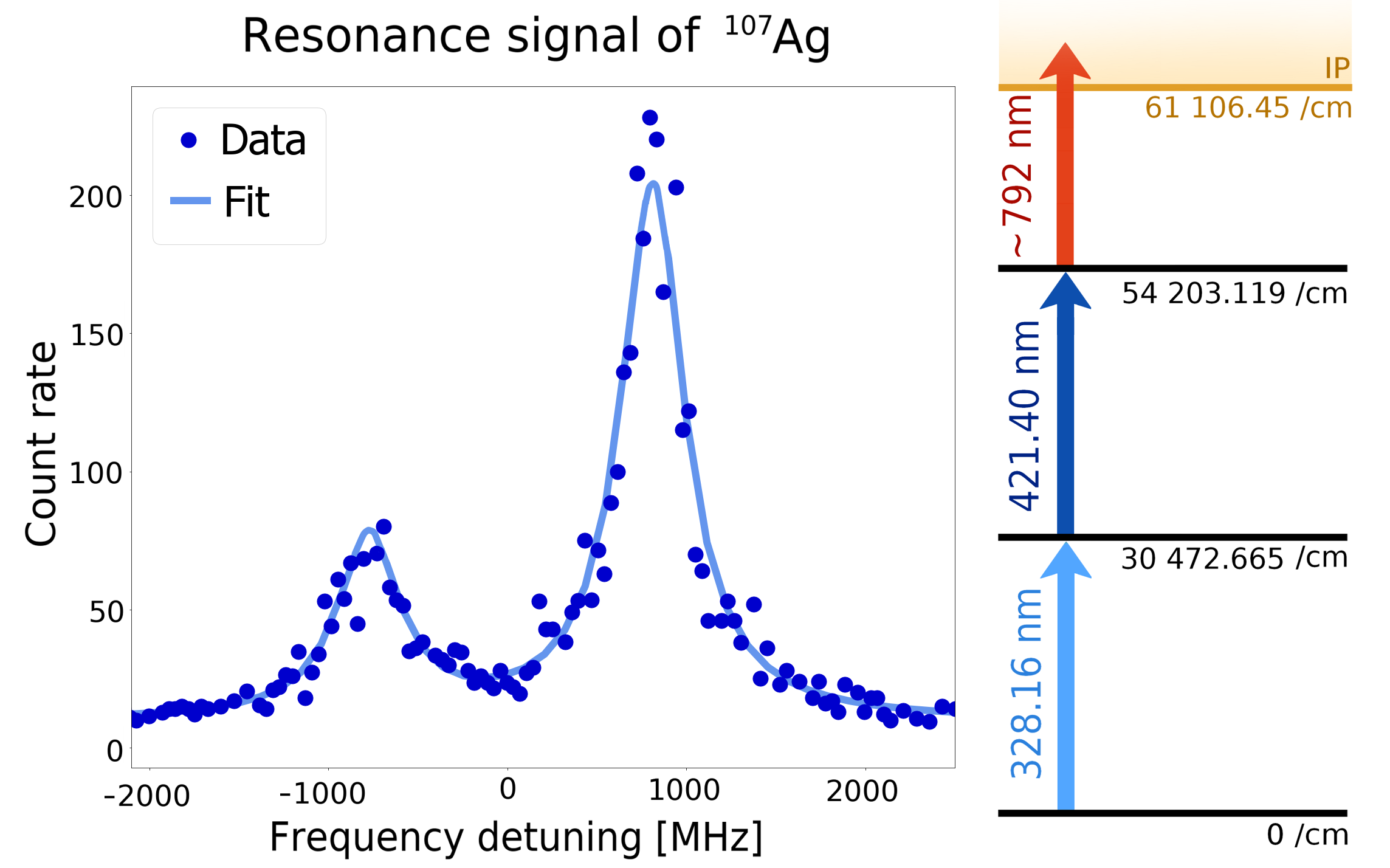}
    \caption{A high-resolution laser resonance scan of the first-step transition in stable $^{107}$Ag using RAPTOR, obtained using the ionization scheme pictured on the right. The notation IP is the ionization potential.}
    \label{fig:hfs}
\end{figure}

High-resolution resonance ionization signals were recently obtained for stable isotopes of $^{107,109}$Ag, also with 5-keV beam energy. An example spectrum of $^{107}$Ag is shown in Fig.~\ref{fig:hfs}. A three-step laser resonance ionization scheme was used, based on the scheme in Ref. \cite{reponen}, exciting the atom from the ground state to an excited state at 30472.665 cm$^{-1}$ with 328.163-nm (in vacuum) laser light, followed by a second resonant transition at 421.399 nm to a state at 54203.119 cm$^{-1}$. Finally, a non-resonant infrared ionization step was used with laser light at a wavelength of $\sim$792 nm. The laser light for the first step was produced using a frequency-tripled injection-seeded pulsed Ti:Sa laser with a bandwidth of $\sim$20 MHz which was seeded by a Matisse Ti-Sa laser operating at 984 nm. Fig.~\ref{fig:hfs}~also shows a preliminary fit of the hyperfine structure, using a Lorentzian lineshape, from which a Full-Width-at-Half-Maximum of $\sim$ 400 MHz can be extracted. Further optimization of the linewidth and an investigation of the contributions of the laser bandwidth, power broadening, and energy spread of the ion beam will be reported elsewhere. Additionally, a quantification of the laser ionization efficiency and collisional ionization processes will be performed, along with  the transport of ion beams to the JYFLTRAP beamline.

\section{Conclusions}
\label{sec:conc}

RAPTOR, a new CRIS-type beam line has been constructed and partially commissioned at the IGISOL facility, University of Jyväskylä. Ion-beam simulations have been performed to benchmark the tuning voltages for the different electrodes. Stable ion beams have been delivered through RAPTOR indicating the simulated voltages agree within 20\% of the optimized values. Further optimization of the beamline is expected to improve the transmission, and future tests will include coupling of the beams to the JYFLTRAP Penning trap beamline. The charge-exchange cell has been tested on stable ion beams of Cu, Ag and Sn, with neutralization efficiencies of a few 10\% achieved. The first high-resolution resonance ionization spectrsa have been obtained using stable isotopes of silver, using a frequency-tripled injection-locked Ti:Sa laser for the first step transition. The spectral linewidth is estimated as $\sim$ 400 MHz, marking an important milestone for RAPTOR. 

Prior to the first online radioactive ion beam experiment, further efforts to characterize the efficiency of the resonance ionization process are required. Several scientific cases for first experiments have been identified, including a study of short-lived isomeric states of bismuth to explore the interplay of nuclear pairing and deformation on nuclear charge radii, and isomeric-state purification of neutron-rich silver isotopes for trap-assisted decay spectroscopy. RAPTOR will also offer the opportunity to explore the high-precision frontier via laser-RF double-resonance spectroscopy, for example the study of hyperfine anomalies and magnetic octupole moments. In particular indium and bismuth offer large sensitivity to the latter effect (\cite{bi1, bi2}), and are the first candidates for such studies.

\section{Acknowledgements}
\label{sec:ack}
S.K. acknowledges financial support from the Vilho, Yrjö and Kalle Väisälä Foundation of the Finnish Academy of Science and Letters.
A.R. and J.W. received funding from the European Union’s Horizon 2020 research and innovation programme under grant agreement no. 861198–LISA–H2020-MSCA-ITN-2019.

 \bibliographystyle{elsarticle-num} 
 \bibliography{cas-refs}





\end{document}